\documentclass[]{agujournal2019}
\usepackage{url}

\usepackage{subfig}
\usepackage{lineno}
\usepackage[margins]{trackchanges} 
\usepackage{soul}
\usepackage{graphicx}
\usepackage{float}
\usepackage{amssymb, amsmath}
\graphicspath{{Figs/}}

\draftfalse

\addeditor{AH}

\usepackage{todonotes}

\journalname{Space Weather}
\draftfalse

\begin{document}

%
%

\title{Probabilistic prediction of Dst storms one-day-ahead using Full-Disk SoHO Images}

%
%




\authors{A. Hu\affil{1, 2}, C. Shneider\affil{1}, A. Tiwari\affil{1}, E. Camporeale\affil{2,3}}

\affiliation{1}{Centrum Wiskunde \& Informatica, Amsterdam, The Netherlands}
\affiliation{2}{CIRES, University of Colorado, Boulder, CO, USA}
\affiliation{3}{NOAA Space Weather Prediction Center, Boulder, CO, USA}




\correspondingauthor{A. Hu}{andong.hu@colorado.edu; huan.winter@gmail.com}




\begin{keypoints}
\item A new $Dst$ probability model developed from SoHO images using Convolutional Neural Networks (CNNs) with a least-squares based ensemble technique is proposed.
\item The proposed model can well forecast $Dst$ probability at least 1 day ahead during strong storm periods.  
\item The proposed model can capture the signature of strong storm events from SoHO images.

\end{keypoints}

%
%

%
%


\begin{abstract}

We present a new model for the probability that the Disturbance storm time (Dst) index exceeds -100 nT, with a lead time between 1 and 3 days. $Dst$ provides essential information about the strength of the ring current around the Earth caused by the protons and electrons from the solar wind, and it is routinely used as a proxy for geomagnetic storms. The model is developed using an ensemble of Convolutional Neural Networks (CNNs) that are trained using SoHO images (MDI, EIT and LASCO). The relationship between the SoHO images and the solar wind has been investigated by many researchers, but these studies have not explicitly considered using SoHO images to predict the $Dst$ index.

This work presents a novel methodology to train the individual models and to learn the optimal ensemble weights iteratively, by using a customized class-balanced mean square error (CB-MSE) loss function tied to a least-squares (LS) based ensemble.

The proposed model can predict the probability that $Dst<-100 $nT 24 hours ahead with a True Skill Statistic (TSS) of 0.62 and Matthews Correlation Coefficient (MCC) of 0.37. The weighted TSS and MCC from \citeA{guastavino2021bad} is 0.68 and 0.47, respectively. An additional validation during non-Earth-directed CME periods is also conducted which yields a good TSS and MCC score. 
\end{abstract}

\section*{Plain Language Summary}
Geomagnetic storms pose one of the most severe space weather risks to our space borne and ground-based electronic instruments, such as GNSS and radio transmission systems. Dst is one of the most accurate geomagnetic storm indicators. Hence, those storm can be predictable if Dst can be forecasted. Currently, the best Dst model can only predict Dst in several hours. In this study, we present a machine learning based ensemble method to predict the Dst 1-3 days in advance from solar images.
%
%

%


%
%
%
%

\section{Introduction}
\label{sec:introduction}

Geomagnetic storms pose one of the most severe space weather risks to our space-borne and ground-based electronic instruments, such as GNSS and radio transmission systems. A geomagnetic storm can be indicated by several geomagnetic indices such as Kp, ap, and the Disturbance storm time ($Dst$) index \cite{rostoker1972geomagnetic}. These indices are related to the perturbation of the geomagnetic field as measured on local regions on Earth at middle, high, and low latitudes, respectively. Although it is now recognized that a single index is not able to capture and define all geospace storms \cite{borovsky2017dst}, they are routinely used by space weather operational agencies as proxies for geomagnetic activity (see, e.g. https://www.swpc.noaa.gov/products/geospace-geomagnetic-activity-plot). Here, we focus specifically on $Dst$, given the large amount of literature devoted to its prediction, notably using data-driven and machine learning techniques \cite{camporeale2019challenge}. $Dst$ is understood to be a proxy for ring current density \cite{liemohn2001dominant}and it is currently defined by using quasi real-time geomagnetic field measurements from four equatorial ground magnetometer stations: Hermanus, Honolulu, San Juan and Kakioka \cite{sugiura1991equatorial}. 

Most of the current models predict $Dst$ based on solar wind parameters such as the North-South component of the interplanetary magnetic field (IMF) $Bz$ \cite{saiz2008forecasting}. Neural networks have been widely used in modeling $Dst$ empirically. \citeA{lundstedt2002operational} was one of the first to implement a multi-layer perception (MLP) neural network based on IMF $Bz$, solar wind density and velocity, in order to forecast $Dst$ 1-hr in advance. \citeA{saiz2008forecasting, bala2012improvements, lazzus2017forecasting} presented models to forecast $Dst$ up to 6 hours in advance. A Gaussian Process model has been introduced by \citeA{chandorkar2017probabilistic} and \citeA{chandorkar2018probabilistic} and later combined with a a long short-term memory (LSTM) architecture in \citeA{gruet2018multiple} to provide a probabilistic forecast up to 6 hours in advance. An ensemble learning algorithm has been used in \citeA{xu2020prediction}.
\citeA{laperre2020dynamic} evaluated the performance of a LSTM model based on a Dynamical Time Warping (DTW) method. 

All the empirical models mentioned above are trained from solar wind data which are measured in quasi real-time by the ACE or DSCOVR satellites orbiting around the first Lagrangian point (L1), or alternatively using the NASA OMNI database (\url{https://omniweb.gsfc.nasa.gov/}). Hence, in an operational setting, their lead-time would be limited to only a few-hours ahead. 

In this paper, we aim to predict $Dst$ with a longer lead time (in the range of 1 to 3 days ahead) using solar images from the Solar and Heliospheric Observatory (SoHO) as inputs. SoHO is a joint mission between the National Aeronautics and Space Administration (NASA) and the European Space Agency (ESA) and was the first space-based telescope to serve as an early warning system for space weather. Solar images can be observed by a suite of on-board instruments on SoHO \cite{SoHO_overview}, including the Michelson Doppler Imager \cite{SoHO_mdi} (MDI) for the solar photosphere, the Extreme ultraviolet Imaging Telescope \cite{SoHO_eit} (EIT) for the stellar atmosphere to low corona, and the Large Angle and Spectrometric Coronagraph \cite{SoHO_lasco} (LASCO). Although used much less than solar wind data for forecasting purposes, it is known that a significant correlation exists between EIT and $Dst$ index. A semi-physical model called `Anemomilos' is then developed based on this relationship to predict $Dst$ index in 6 days. \cite{tobiska2013anemomilos} This model became part of the US
Space Force HASDM predictions in 2012. \citeA{upendran2020solar} has pointed out that the correlations between solar images and solar wind parameters are most significant during the fast solar wind. 

Obviously, by setting the problem as a 1-to-3 days ahead forecast, we have to accept that we cannot achieve the accuracy seen in few (1 to 6) hours ahead forecast models, that currently report Root Mean Square Errors of the order of 10 nT or less. Therefore, as a first step, we set the problem as a classification task, aiming at forecasting the probability that $Dst$ exceeds a certain threshold (hereinafter referred to as `$Dst$ probability'). 
In this study, we focus on strong storms having a $Dst$ threshold of -100 nT, and aiming at producing a probabilistic forecast 1 day ahead of a given solar image.

In addition, although most operational
applications require a deterministic Dst value, the predicted probabilistic Dst forecast could also be used to improve current space weather models, for example, for running ensemble simulations of mass density forecasting, which is one of the top priority for the predictability of low-Earth-orbit (LEO) satellite trajectories \cite{licata2020benchmarking}. $Dst$  plays a major role in mass density modeling such as Jacchia-Bowman2008 (JB2008)\cite{bowman2008new}. Hence, The forecast $Dst$ probability would be helpful to assess the uncertainty of those models.


We train a machine learning (ML) technique called convolutional neural network (CNN) to forecast the probability that $Dst$ exceeds the pre-defined -100 nT threshold from 1 to 3 days in advance (i.e., the prediction is in the form of a time series of probabilities). CNN has been recently used in space weather applications, eg. by \citeA{siciliano2021forecasting, upendran2020solar,li2020predicting,ruwali2020implementation,park2018application}. 
By using the presented technique in an operational setting, a forecaster would have access to several predictions issued with different lead-times. Hence, we face a classical problem in ensemble learning, namely how to combine different predictions by applying different weights to different lead-times. In this work, we restrict to a static weighting scheme (i.e. the weights are learned on a training set and do not change with different inputs or solar wind conditions), opposite to dynamic weights \cite{polikar2012ensemble}. We solve the ensemble problem by introducing a new, customized, complementary cumulative distribution function (CCCDF) based least-squares (LS) method to find the optimal weights.

The paper is divided as follows. Section \ref{sec:Data} introduces the data used for this study, the criterion to select storm times and the corresponding time periods covered. Section \ref{sec:methods} describes the methodology, including the designed machine learning architecture, the optimization method, and the performance metrics for assessment. Section \ref{sec:results} presents the results of the developed model, and emphasizes the probabilistic nature of the forecast. Finally, in Section \ref{sec:summary-outlook}, we draw conclusions and make final remarks about future directions.

\section{Data}
\label{sec:Data}

\subsection{Disturbance storm time ($Dst$) index}
\label{subsec:Dst}

\begin{figure}
  \centering
  \includegraphics[width=\textwidth]{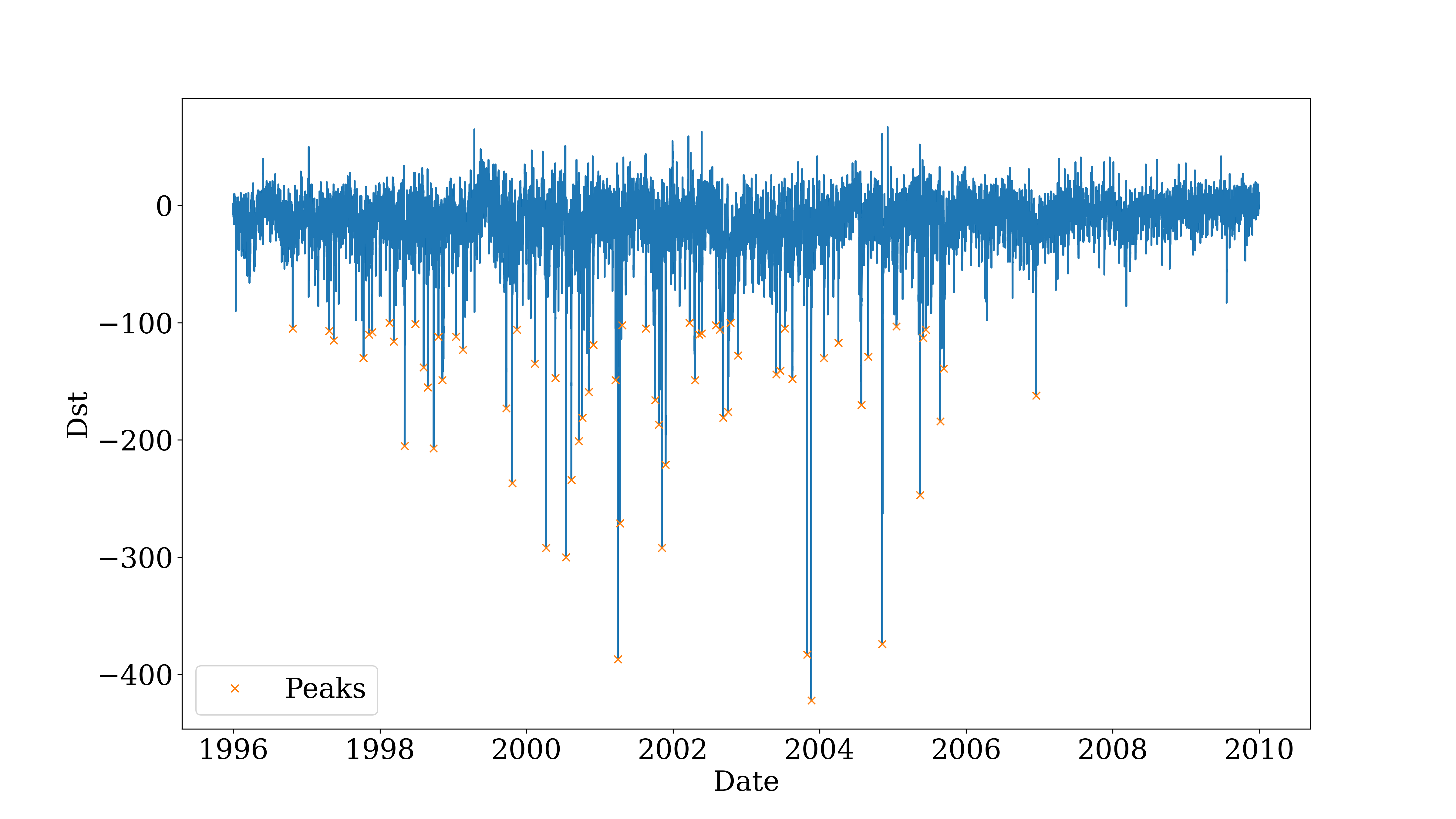}
  \caption{Time history of $Dst$ during 1996 to 2010. X axis is date and Y axis is $Dst$ value. The orange crosses denote peak values smaller than -100 nT, used for defining storm events considered in this study.}
  \label{fig:Peak_select}
\end{figure}

\begin{figure}
  \centering
  \includegraphics[width=0.7\textwidth]{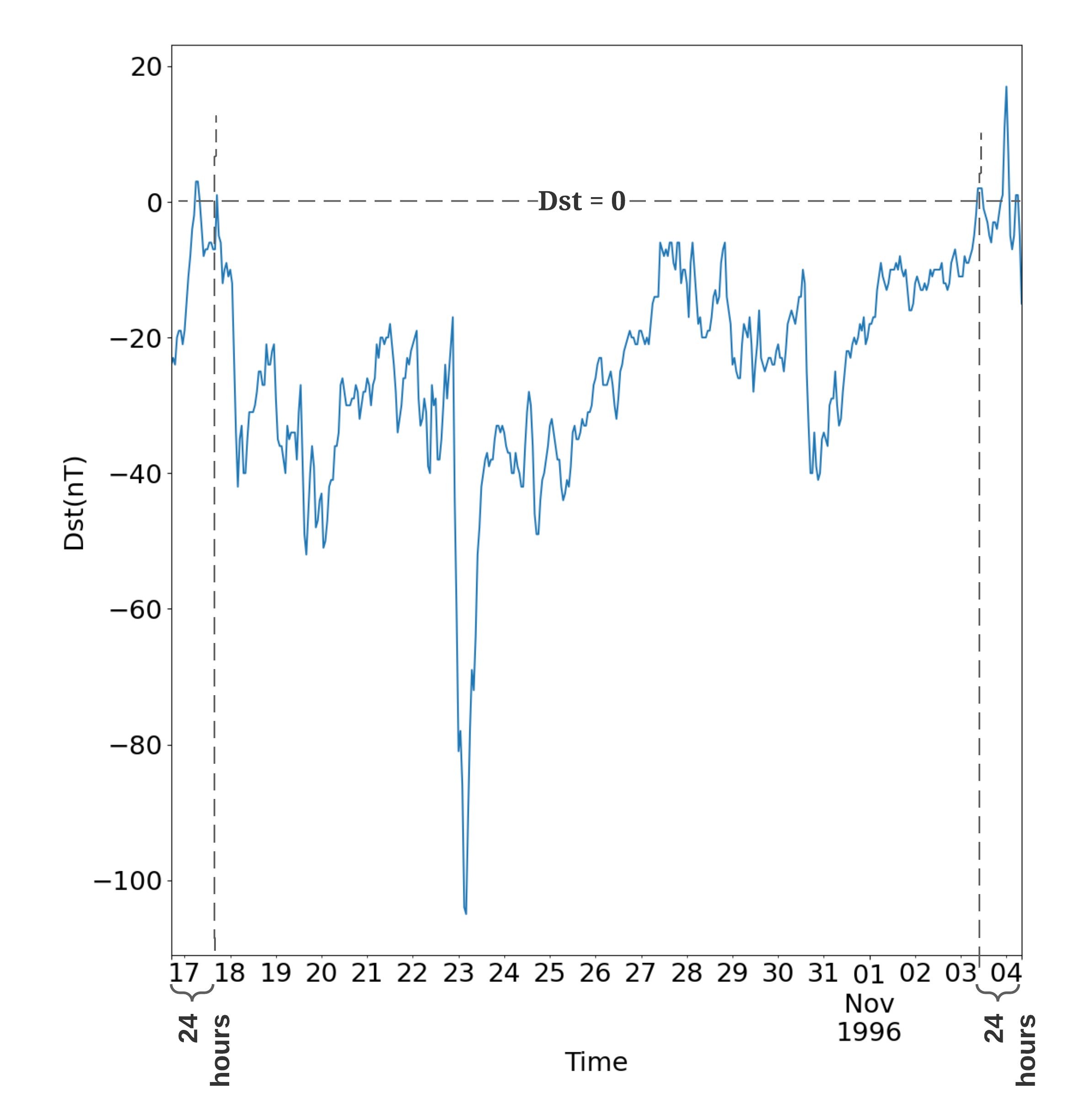}
  \caption{An example of the selection criterion to define the time range for one storm event. The $Dst$ peak occurs on Oct. 23, 1996. The nearest positive $Dst$ values before and after the peak occur on Oct. 18  and Nov. 03, respectively. The whole storm range is defined between Oct. 17, 1996 and Nov. 04, 1996 with a 24-hour buffer zone. The list of selected storm events can be found in Table. \ref{tab:storms}.}
  \label{fig:range_select}
\end{figure}

The $Dst$ index is available at 1-hour cadence from the NASA OMNI database. Fig. \ref{fig:Peak_select} displays the $Dst$ index in the period 1996-2010. The model is trained, validated and tested on storm events with a $Dst$ peak smaller than -100 nT, shown by orange crosses. Overall, 51 such storm periods are selected for this study. In order to define a storm period, we look for the the nearest positive $Dst$ values immediately before and after each peak, and then extend the time window by a 24-hour buffer zone to make sure that the pre-storm period and the recovery phase are fully included. An example is shown in Fig. \ref{fig:range_select}, where the Dst peak is observed on Oct. 23, 1996. The storm period is defined as ranging between Oct. 17, 1996 and Nov. 04, 1996. With this procedure we make sure that the time intervals are selected in such a way that the negative $Dst$ peaks do not always occur at the same time within the chosen storm-time window, hence the neural network does not simply memorize. The average period of selected storm events is approximately 15 days. All selected storms, sorted by peak $Dst$, are listed in Table. \ref{tab:storms}.  

\begin{table}[!htbp]
\centering
\caption{List of selected 51 Storm Events.}
\begin{tabular}{c c c c}
  \hline
  No. & Start time & End time & Min. Dst (nT)\\
  \hline
1 & 2001-03-29 03:00:00 & 2001-04-06 15:00:00 & -387 \\
2 & 2001-11-03 19:00:00 & 2001-11-17 00:00:00 & -292 \\
3 & 2005-05-13 05:00:00 & 2005-05-22 04:00:00 & -247 \\
4 & 1999-10-19 23:00:00 & 1999-11-02 11:00:00 & -237 \\
5 & 2000-08-08 05:00:00 & 2000-08-21 07:00:00 & -234 \\
6 & 2001-11-22 06:00:00 & 2001-12-02 14:00:00 & -221 \\
7 & 2000-09-15 19:00:00 & 2000-09-26 14:00:00 & -201 \\
8 & 2001-10-17 10:00:00 & 2001-10-27 09:00:00 & -187 \\
9 & 2005-08-22 08:00:00 & 2005-09-02 03:00:00 & -184 \\
10 & 2002-09-01 23:00:00 & 2002-09-18 07:00:00 & -181 \\
11 & 1999-09-20 20:00:00 & 1999-09-28 15:00:00 & -173 \\
12 & 2000-11-02 04:00:00 & 2000-11-12 06:00:00 & -159 \\
13 & 2001-03-17 11:00:00 & 2001-03-24 14:00:00 & -149 \\
14 & 2003-08-15 18:00:00 & 2003-09-02 23:00:00 & -148 \\
15 & 2003-06-14 09:00:00 & 2003-06-28 12:00:00 & -141 \\
16 & 2000-02-09 07:00:00 & 2000-02-21 12:00:00 & -135 \\
17 & 2004-01-20 05:00:00 & 2004-02-01 05:00:00 & -130 \\
18 & 2004-08-28 02:00:00 & 2004-09-06 01:00:00 & -129 \\
19 & 2000-11-24 22:00:00 & 2000-12-05 05:00:00 & -119 \\
20 & 2002-05-09 11:00:00 & 2002-05-20 20:00:00 & -110 \\
21 & 2002-05-21 11:00:00 & 2002-06-01 20:00:00 & -109 \\
22 & 2002-08-16 22:00:00 & 2002-08-27 10:00:00 & -106 \\
23 & 2001-08-15 16:00:00 & 2001-08-21 12:00:00 & -105 \\
24 & 2005-01-14 21:00:00 & 2005-01-23 17:00:00 & -103 \\
25 & 2002-07-30 23:00:00 & 2002-08-09 03:00:00 & -102 \\
26 & 2002-03-21 15:00:00 & 2002-03-30 21:00:00 & -100 \\
27 & 2000-01-20 16:00:00 & 2000-01-29 06:00:00 & -96 \\
28 & 2005-01-05 14:00:00 & 2005-01-13 18:00:00 & -93 \\
29 & 2004-02-09 10:00:00 & 2004-02-23 20:00:00 & -93 \\
30 & 1999-04-14 20:00:00 & 1999-04-23 08:00:00 & -91 \\
31 & 2000-06-06 13:00:00 & 2000-06-15 01:00:00 & -90 \\
32 & 2002-01-31 00:00:00 & 2002-02-06 21:00:00 & -86 \\
33 & 1999-12-02 02:00:00 & 1999-12-17 07:00:00 & -85 \\
34 & 2003-05-07 08:00:00 & 2003-05-20 06:00:00 & -84 \\
35 & 2009-07-20 01:00:00 & 2009-08-01 10:00:00 & -83 \\
36 & 2010-03-29 23:00:00 & 2010-04-16 16:00:00 & -81 \\
37 & 2005-02-14 12:00:00 & 2005-02-24 09:00:00 & -80 \\
38 & 2000-01-09 10:00:00 & 2000-01-19 11:00:00 & -80 \\
39 & 2010-05-26 21:00:00 & 2010-06-10 15:00:00 & -80 \\
40 & 2004-03-07 13:00:00 & 2004-03-21 21:00:00 & -78 \\
41 & 2004-07-14 22:00:00 & 2004-07-22 18:00:00 & -76 \\
42 & 2001-05-05 00:00:00 & 2001-05-18 18:00:00 & -76 \\
43 & 2000-06-24 02:00:00 & 2000-07-02 09:00:00 & -75 \\
44 & 2002-12-17 08:00:00 & 2002-12-24 16:00:00 & -75 \\
45 & 2005-10-29 09:00:00 & 2005-11-04 18:00:00 & -74 \\
46 & 2003-01-27 13:00:00 & 2003-02-15 13:00:00 & -74 \\
47 & 2007-03-21 10:00:00 & 2007-03-27 00:00:00 & -72 \\
48 & 2002-02-26 17:00:00 & 2002-03-05 01:00:00 & -71 \\
49 & 2010-04-30 11:00:00 & 2010-05-12 17:00:00 & -71 \\
50 & 1999-10-08 01:00:00 & 1999-10-22 04:00:00 & -67 \\
51 & 2003-02-24 22:00:00 & 2003-03-14 15:00:00 & -67 
\end{tabular}
\label{tab:storms}
\end{table}

\begin{figure}
  \centering
  \includegraphics[width=.8\textwidth]{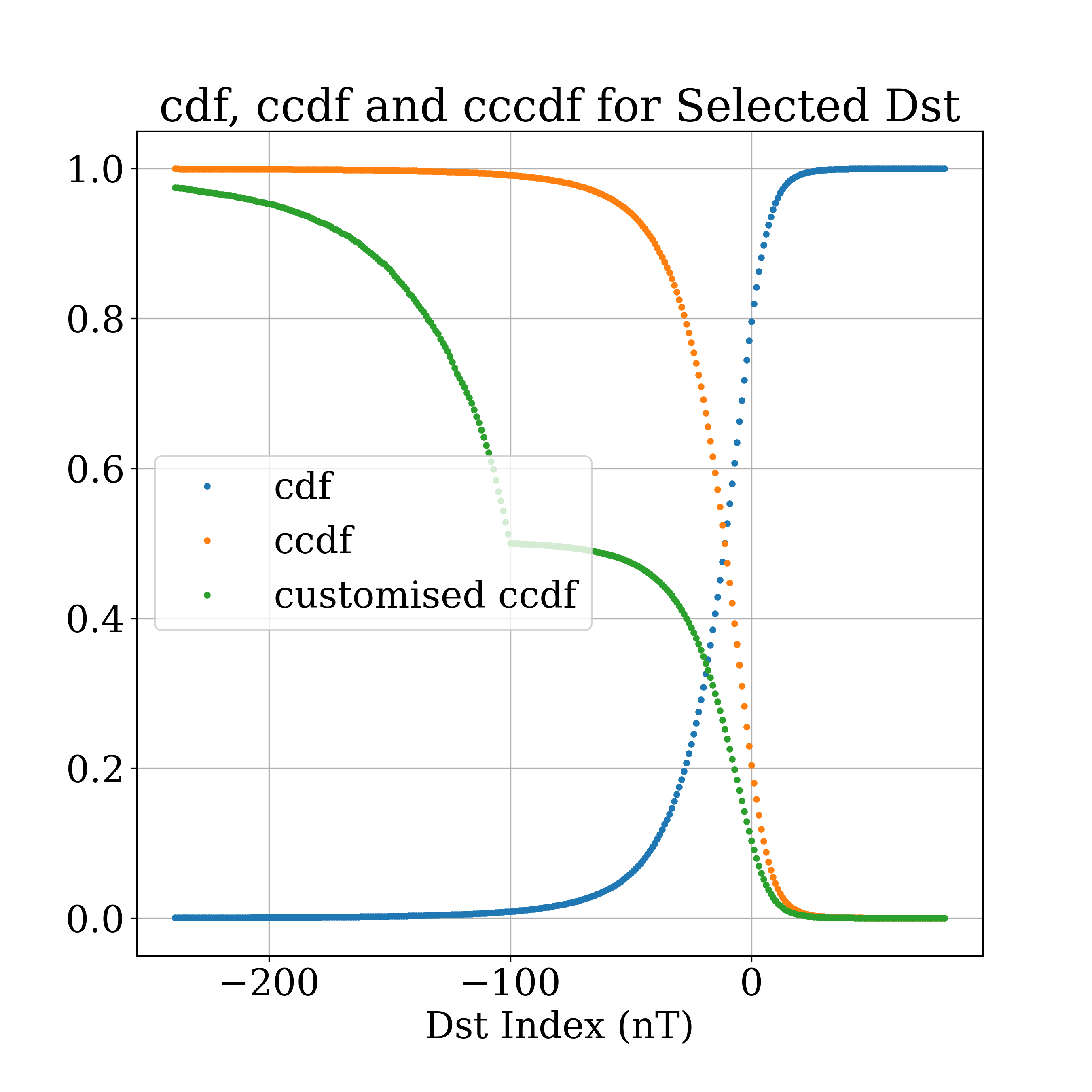}
  \caption{CDF, CCDF and customized CCDF of $Dst$ during storm periods. X axis is $Dst$ index. Blue dots are CDF; Orange dots are CCDF; and green dots are CCCDF in this study.}
  \label{fig:pdf_cdf}
\end{figure}

As mentioned in the Introduction, we would like to solve this classification task by a regression model. Hence, instead of a binary label set (positive/negative), a customized complementary cumulative probability distribution function (CCCDF) of $Dst$ is used as a target for the CNN model. The CCCDF is shown in Fig. \ref{fig:pdf_cdf}, and explained briefly below. A cumulative distribution function (CDF) is defined as the integral of a probability density function (PDF) from negative infinity to $x$ and the complementary CDF (CCDF) = 1-CDF or the integral from positive infinity to $x$. The CDF($x$) is the probability that a random variable has a value less than $x$. Conversely, the CCDF($x$) gives the probability that the variable under consideration is larger than $x$. The customized CCDF (CCCDF) for $Dst$=-100nT is defined as in Eqn. \ref{eqn:pos}-\ref{eqn:neg}. Note that CCCDF($Dst$=-100) =0.5 by construction. 
\begin{equation}
\label{eqn:pos}
CCCDF(x) = \frac{(CCDF(x) - CCDF(-100))}{1-CCDF(-100)}+0.5  \mbox{      for } x \le -100
\end{equation}

\begin{equation}
\label{eqn:neg}
CCCDF(x) = \frac{(CCDF(x) - CCDF(-100))}{CCDF(-100)}+0.5 \mbox{       for }  x > -100
\end{equation}

\subsection{SoHO mission}
\label{subsec:SoHo}


The two-hourly SoHO data sets used in this work for the period 1996-05-01 to 2011-04-12 are derived from the following public domain resources: the NASA Solar Data Analysis Center’s (SDAC) Virtual Solar Observatory (VSO) (\url{https://sdac.virtualsolar.org/cgi/search}) and Stanford University’s Joint Science Operation Center (JSOC) \url{http://jsoc.stanford.edu/MDI/MDI_Magnetograms.html}. All SoHO products and their details are shown in Table 2.

More than 20,000 SOHO images can be provided from three on-board instruments, including the Michelson Doppler Imager (MDI) for the solar photosphere, the Extreme ultraviolet Imaging Telescope (EIT) for the stellar atmosphere to low corona, and the Large Angle and Spectrometric Coronagraph (LASCO) covering the corona from $1.5-30~R_s$. Those data have fully covered Solar Cycle $23$ and $24$. Among them, MDI, EIT with a wavelength of 195 (EIT-195) and LASCO-C2 are used as the inputs of this study.

\begin{figure}[h!] 
\centering
\includegraphics[width=1\textwidth]{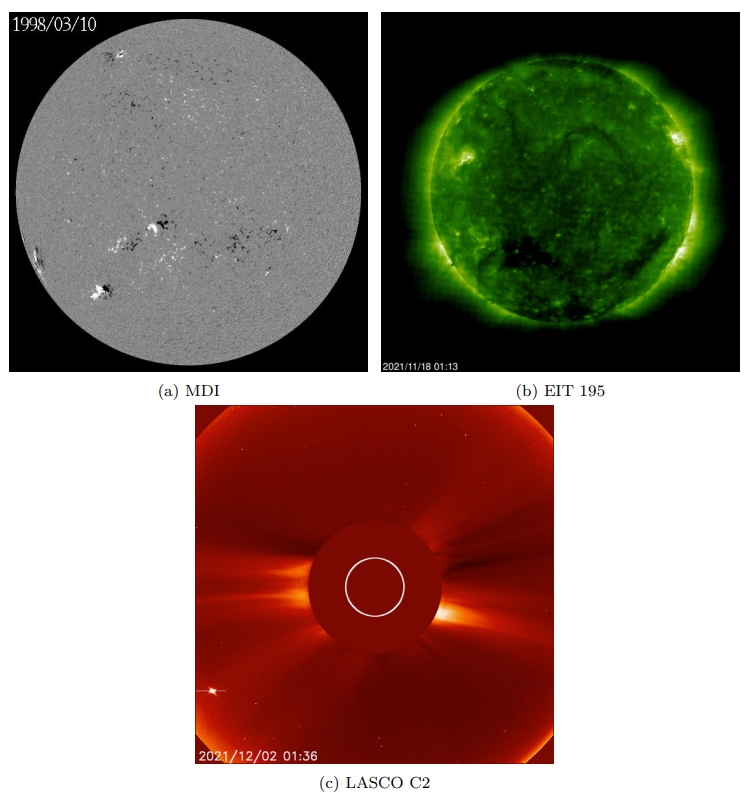} 
\caption{Examples of various SoHO products (a-c). Colors similar to those used on the NASA SoHO site have been used.}
\label{fig:holes}
\end{figure}

However, the SDAC data is highly heterogeneous. Not only are there intrinsic differences among these SoHO products (e.g., individual cadence for each channel shown in Table 2), but there is also an irregular assortment of image file sizes and processing levels. All products require calibration before they can be used for the neural networks. In addition, each channel needs to be synchronized with a fixed cadence (i.e., 2hrs in this study). An example of the calibrated data are shown in Fig. \ref{fig:holes}. 

A pipeline has been created and published by \cite{2021arXiv210806394S} for automatically downloading, cleaning and synchronizing these original images from SDAC and VSO. A machine-learning-ready image data set is then provided which is a valuable resource for the space weather community.

\begin{table}[htbp!]
\centering
\begin{tabular}{l|l|l|l|l|l}
\hline
Instrument & Detector & Observed Region & $\lambda($\AA$)$ & Cadence (min) & Date Range\\
\hline
MDI & MDI & Full Disk & $6768$ (Ni I) & $\sim 96$ & 1996.05.01 - 2011.04.12 \\
\hline
EIT & EIT & Full Disk & $171$ (Fe IX/X) & $\sim 360$ & 1996.01.01 $\rightarrow$ \\
EIT & EIT & Full Disk & $195$ (Fe XII) & $\sim 12$ & 1996.01.01 $\rightarrow$ \\
EIT & EIT & Full Disk & $284$ (Fe XV) & $\sim 360$ & 1996.01.01 $\rightarrow$ \\
EIT & EIT & Full Disk & $304$ (He II) & $\sim 360$ & 1996.01.01 $\rightarrow$ \\
\hline
LASCO & C2 & Corona ($1.5 - 6~R_s$) & Visible & $\sim 20$ & 1995.12.08 $\rightarrow$ \\
LASCO & C3 & Corona  ($3.5 - 30~R_s$) & Visible & $\sim 20$ & 1995.12.08 $\rightarrow$
\end{tabular}
\caption{Suite of SoHO Instruments utilized. $\lambda($\AA$)$ is wavelength measured in angstroms, and $R_s$ is the Sun's radius. LASCO C1 ($1.1 - 3~R_s$) is not included in this work since it was only operational till Aug. 9, 2000.} 
\label{tab:SoHO}
\end{table}

\section{Methodology}
\label{sec:methods}

The goal of this study is to estimate the $Dst$ probability 1 to 3 days ahead of the time when full-disk SoHO images are taken. For the sake of clarity we discuss here the algorithm for 1-day ahead prediction, with the understanding that all times are correspondingly shifted for 2 and 3 days ahead predictions.

\subsection{Customized Class-Balanced Convolutional Neural Networks (CB-CNN)}
\label{subsec:CNN}
Several machine learning approaches can be used for a probabilistic prediction task. We have compared naive Bayes, multi-layer perception and CNN. Among them, CNN has yielded the most reliable and robust performance (results are not shown here). 

CNN is a commonly used neural network architecture, widely used in computer vision \cite{gu2018recent}, in solar image processing \cite{illarionov2018segmentation,baso2018enhancing,upendran2020solar,dos2021multichannel}, and recently in plasma and space physics applications \cite{hu2020identifying,siciliano2021forecasting}. Hence, in this study, we have opted to use a customized class-balanced convolutional neural network (CNN). This is because of the imbalance between the number of positive samples ($Dst<=-100 nT$) and the number of negative samples ($Dst>-100 nT$), which is approximately 10\% of all training samples. As a result of using a class-balanced target, the CB-CNN model performs better than a vanilla CNN model, when the target $Dst$ is near the classification threshold. A brief introduction of the CNN architecture and the optimization methods used in the training is listed in Table \ref{tab:CNN_architecture}.

\begin{table}[!htbp]
    \centering
    \begin{tabular}{c|c}
         Parameter & Value \\
         \hline
         \textbf{input size} & \textbf{[3, 256, 256]} \\
         \textbf{output size} & \textbf{12} \\
         layers & 3 \\
         kernel size & [3, 5, 5] \\
         \textbf{padding mode} & \textbf{same} \\
         activation function & [ReLU, ReLU, ReLU]\\
         max epochs & 100\\
         optimization & AdamW\\
         learning rate & 0.0001\\
         regularization & Elastic \\
         cost function & CB-MSE 
    \end{tabular}
    \caption{Parameter selection of CNN and the corresponding optimization method used in training.}
    \label{tab:CNN_architecture}
\end{table}

By denoting with $T$ the time at which the input images are taken, a multi-target CNN model is first trained to predict the $Dst$ probability in the time range $[T+26,T+48]$, with a time resolution of 2 hours. That is, 12 probability values are output for each input. This means that, at any given time, we have 12 probabilities that have been predicted between 26 and 48 hours ahead. Note that, due to the variability of the Sun and because we are using full disk images, predictions with shorter time lags are not necessarily more accurate than ones with longer time lags. Because we eventually want to merge those predictions (with different time lags) into a unique, reliable prediction, we do not use a standard loss function for binary classification (such as, e.g., binary cross-entropy), but a customized, class-balanced, mean square error.

In our application, a class-balanced loss function developed by \citeA{cui2019class} is used to deal with the large imbalance between positive and negative labels. In addition to that, we want to penalize more the incorrect predictions that are closes to the decision boundary $CCCDF=0.5$ (corresponding to the threshold $Dst=-100$). Hence, a customized weight is designed to artificially increase the cost function for the possibly ambiguous samples with $Dst$ near -100 nT. The cost function is defined as:

\begin{align}
  \label{eqn:CB}
  \centering
  \mathrm{Cost\ Function =} \sum_{h=26}^{48}w_h\Bigg(\frac{\sum_{i=1}^{n_{pos}^h}E_i^h(1-\beta)}{1-\beta^{n_{pos}^h}}+\frac{\sum_{i=1}^{n_{neg}^h}E_i^h(1-\beta)}{1-\beta^{n_{neg}^h}}\Bigg)
\end{align}

\begin{align}
  \centering
  \label{eqn:En}
  E_i^h=(\hat{P}_i^h-P_i^h)\cos{((P_i^h-0.5)\times 0.9\times\pi}^2
\end{align}

\begin{figure}
    \centering
    \includegraphics[width=1.2\textwidth]{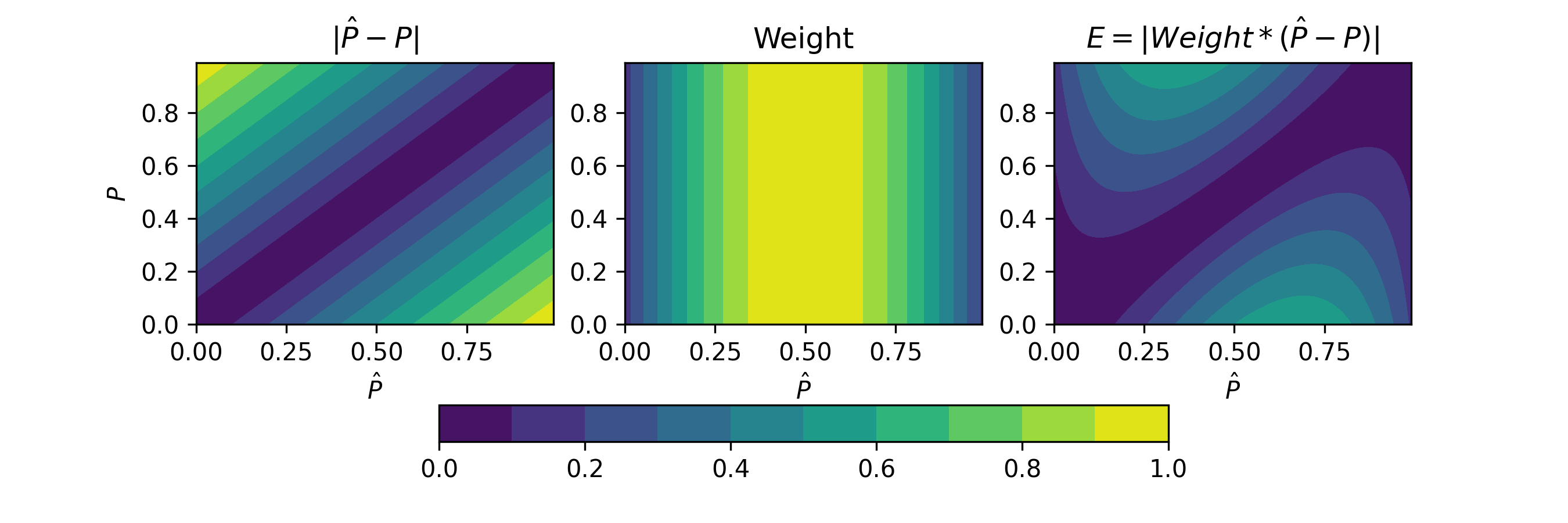}
    \caption{Contour map of Eqn. \ref{eqn:En}. Left panel is absolute error between $\hat{P}$ and $P$; middle panel is the weight in $E_i$ which is the second term in Eqn. \ref{eqn:En}; and the last panel is final $E_i$.}
    \label{fig:Ei}
\end{figure}

Where $P_i^h$ is the model prediction for $i$th sample with a delay hour $h$ (delay hours $h$ ranging 26h, 28h, $\ldots$, 48h), $\hat{P}_i^h$ is the corresponding target, $n_{pos}$ and $n_{neg}$ are the number of positive and negative samples in a batch, respectively, and $w_h$ denotes the weight associated to each target. The detailed procedure of optimizing these weights will be further introduced in Section \ref{subsec:ensemble}. $E$ is a customized square error that penalizes samples whose $Dst$ is near -100 nT, as shown in Fig. \ref{fig:Ei}. 
$\beta$ is a constant term manually set to 0.9999 according to \citeA{cui2019class}.

\begin{figure}
  \centering
  \includegraphics[width=\textwidth]{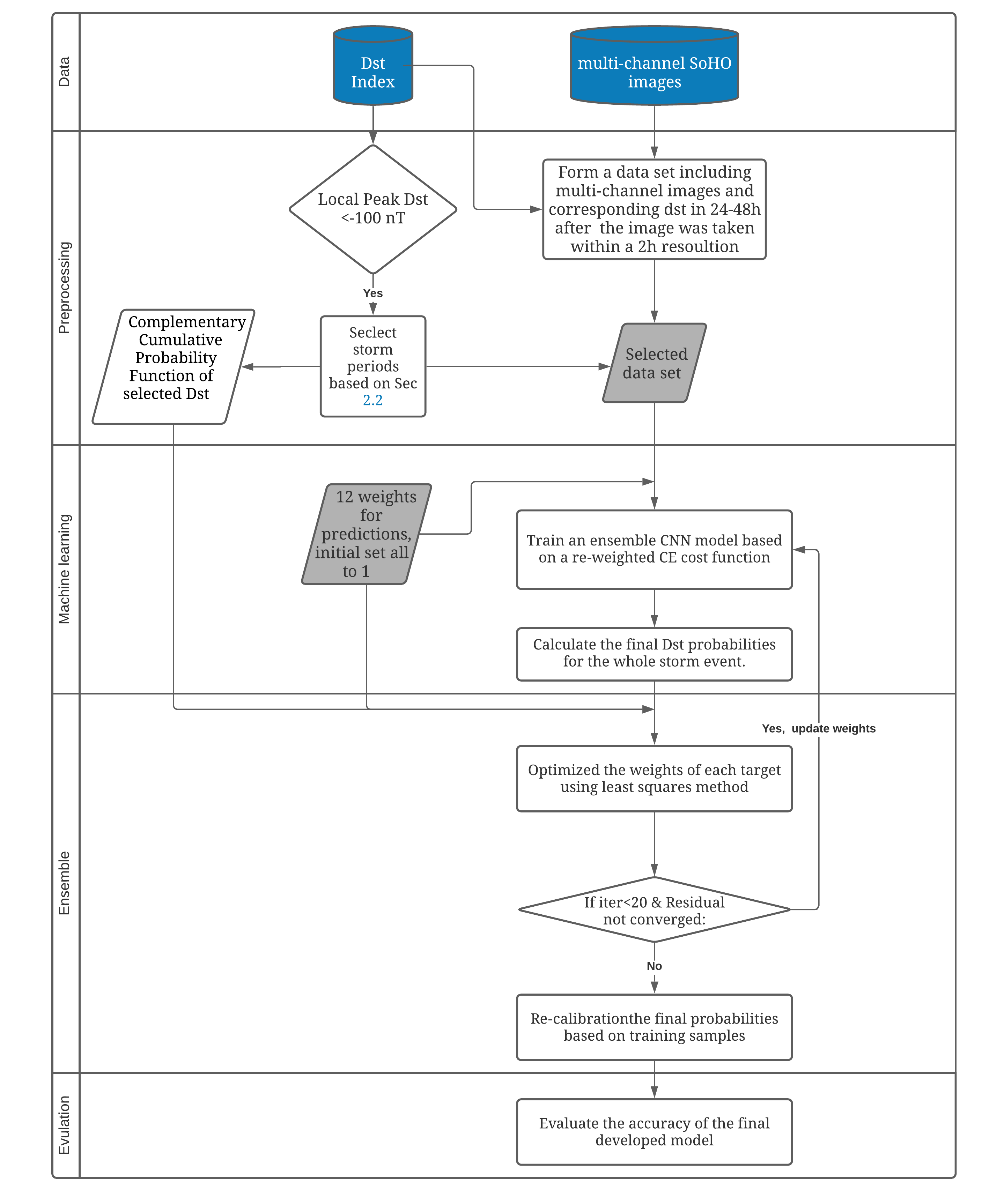}
  \caption{Flowchart of modeling procedures.}
  \label{fig:flowchart}
\end{figure}

The multi-target CNN trained using the cost function described above, can then be used to forecast the Dst probability from 26 to 48 hours ahead, at a 2 hours rate. Hence, at any given time we have $(48 -26)/2+1 = 12$ different predictions (issued between 26 and 48 hours prior). A natural question then arises on whether one could combine these 12 predictions in order to achieve a prediction that is more accurate than any individual one. In order to to that, we take a weighted average of the 12 predictions, and we estimate the optimal (static) weights by solving a least squares (LS) problem. The details of the weight estimate procedure are shown in Fig. \ref{fig:flowchart}. Figure \ref{fig:weight} shows the learned 12 weights optimized with this procedure, along with their uncertainty (see the leave-one-out procedure described below). 

\begin{figure}
  \centering
  \includegraphics[width=0.8\textwidth]{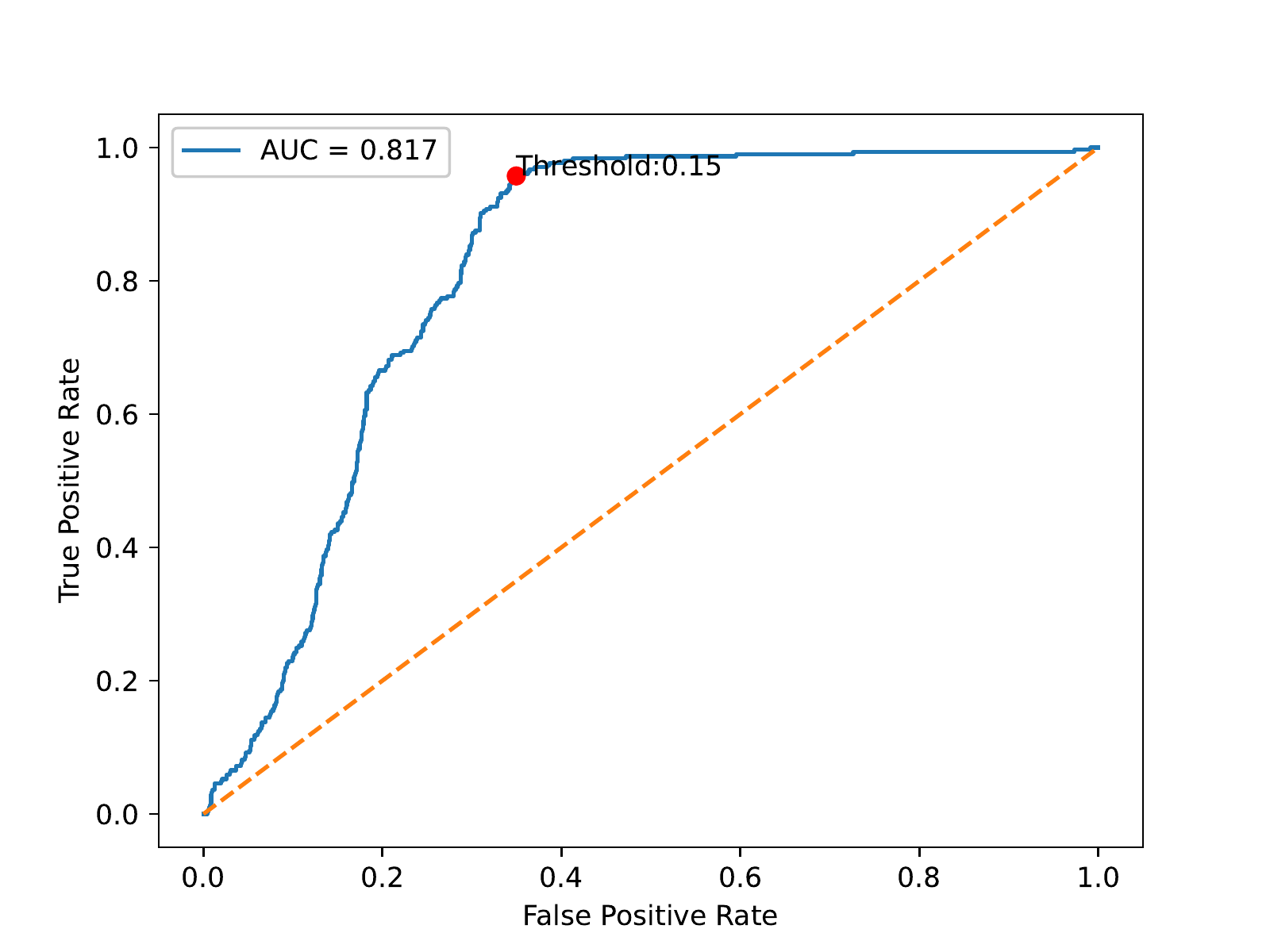}
  \caption{ROC curves (TP rate vs FP rate). X and Y axes are FP rate and TP rate respectively. Red dots indicate the optimal points along this given ROC curve.}
  \label{fig:ROC}
\end{figure}

Finally, we notice that the predicted Dst probabilities might not be well-calibrated (i.e., statistically consistent with observations) and that the optimal threshold for binary classification metrics might be different than the standard 50\% probability. Hence, the threshold of probability used for metrics (see Metrics Section) is re-calibrated by using a receiver operating characteristic (ROC) curve. ROC curve is an important diagnostic for a probabilistic model that can be used to determine an optimal threshold to separate positives from negatives based on probabilistic predictions. A detailed description of ROC curve can be found in \citeA{camporeale2020gray}. An example ROC curve for the developed model with the CNN is shown in Fig. \ref{fig:ROC}. Horizontal and vertical axes denote false positive rate (FPR) and true positive rate (TPR), respectively. The dashed orange line shows TPR equals to FPR (i.e., no skill), while the blue line represents the ROC curve, obtained by defining positives and negatives by progressively changing the probability threshold from 0\% to 100\%. The red dot represents the optimal/largest value of True Skill Statistics, defined as the difference (TPR-FPR).

\subsection{Metrics}
\label{subsec:metrics}

In order to precisely assess the accuracy of a model, it is important that the performance metrics are computed on a test set independent from the training set (so-called hold-out data), hence making sure that the machine learning algorithm does actually learn meaningful patterns and does not merely memorize the training data. A `Leave one out' technique is adopted here. That is a K-fold cross validation taken to its logical extreme, with K equal to N, the number of selected storm cases. That means that the proposed model is trained on all the data except for one storm window and a prediction is made for that left-out storm. The procedure is repeated N times. Finally, the metrics are computed as averages over the N models.
In this study, the top 51 storm windows in the period 1999-01-01 to 2011-04-10 constitute each a fold. Details of those events can be found in SI.
The probabilistic predictions can be transformed to binary labels upon defining a probability threshold. In this way we can use standard metrics for binary classification such as the True Skill Statistic (TSS) and Matthews Correlation Coefficient (MCC)\cite{camporeale2020gray}:

\begin{align}
    \centering
    \text{TSS} &= \text{TPR} - \text{FPR} = \frac{\text{TP}}{\text{TP} + \text{FN}} - \frac{\text{FP}}{\text{FP} + \text{TN}}, \label{eqn:TSS} \\
    \label{eqn:mcc}
    \mathrm{MCC} &= \frac{TP \times TN  - FP \times FN}{\sqrt{(TP + FP) \times (FN + TN) \times (FP + TN) \times (TP + FN)}},
\end{align}

where TP, FP, TN and FN denotes true positive, false positive, true negative and false negative numbers respectively. The MCC score is a reliable statistical rate that produces a high score only if the prediction obtained good results in all of the four confusion matrix categories (TP, FP, TN and FN), proportionally both to the size of positive elements and the size of negative elements in the data set \cite{baldi2000assessing}. TSS is a useful metric that combines both types of information and should be as close as possible to 1. Those metrics have shown some advantages over the F1 score and accuracy in binary classification evaluation \cite{chicco2020advantages}. Moreover, an innovative way to evaluate the model accuracy has been developed by \citeA{guastavino2021bad}. This method assign different weights to FPs that anticipate the occurrence of an actual positive event (i.e. 'almost hit'). The value-weighted MCC and TSS have been proved more appropriate for decision making processes. Hence, these weighted scores are also considered to assess the model accuracy.

\subsection{Ensemble Method}
\label{subsec:ensemble}

\begin{figure}
  \centering
  \includegraphics[width=\textwidth]{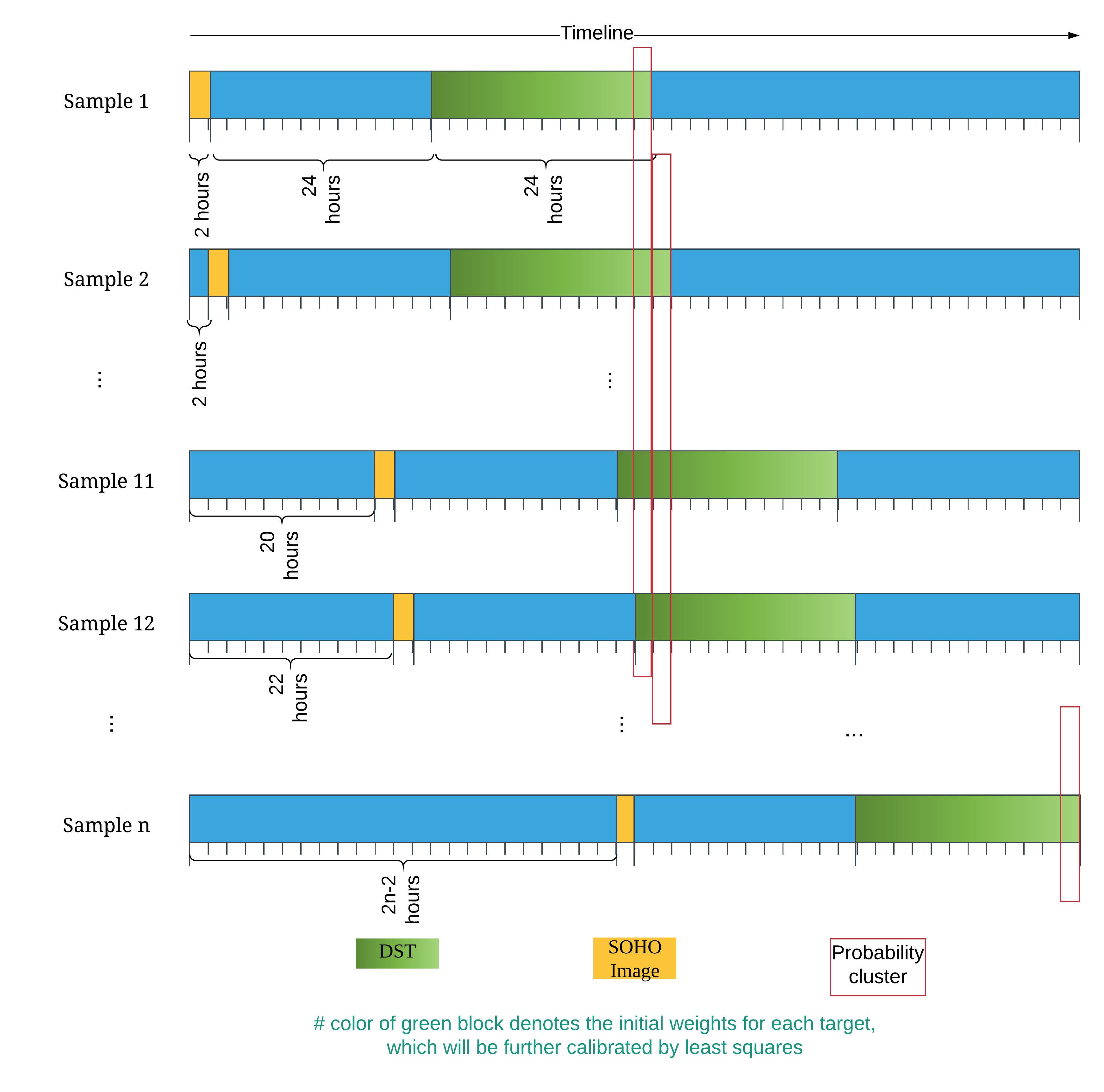}
  \caption{Illustration of the ensemble procedure. Horizontal bar displays the timeline during the whole event. Yellow blocks denote when SoHO images are taken. Green blocks are the 12 $Dst$ probability predictions as introduced from Sec. \ref{subsec:CNN}. The weights of those 12 predictions from each sample are consistent. From top to bottom, there are $n$ samples during this event. The time shift between nearby samples is 2 hours. All predicted probabilities in one red frame is considered as one probability cluster (i.e., a $12\times1$ vector) for a certain time epoch.}
  \label{fig:dst_timeline}
\end{figure}

After the CNN model is developed, twelve probabilities can be predicted from the model and each SOHO image set. Although each prediction is per se valid, we have verified that combining those predictions yields a model that outperforms a single individual prediction. Here, we describe the ensemble method that, for simplicity, has been chosen to be a simple linear combination of the twelve probabilities so that the final probability is defined as $P_{ens}=\sum_{i=26}^{48} w_ip_i$, with $p_i$ the probability of Dst exceeding the -100 nT threshold at time $i$.
The timeline of predictions is depicted in Fig. \ref{fig:dst_timeline}. Each horizontal bar displays the timeline during an hypothetical event. Yellow blocks denote the 2 hours interval during which SoHO images are taken, and used as inputs to the model. The green block denotes the 12 $Dst$ probability predictions. From top to bottom, there are $n$ samples during one event. The time gap between consecutive samples is 2 hours. All predicted probabilities enclosed by a red frame are considered as one probability cluster (i.e., a $12\times1$ vector) for a certain time epoch. The totality of clusters are used to form a design matrix for this event, i.e., a $12\times(n-12)$ matrix. Assuming we have $m$ events, the whole design matrix has then size $12\times \sum_i^m(n_i-12)$. Eqn. \ref{eqn:prob} is the observation model of the ensemble method:
\begin{equation}
\label{eqn:prob}
L=W\cdot P,
\end{equation}

where $P$ is the design matrix as introduced in Fig. \ref{fig:dst_timeline} (each column of $P$ contains 12 different probabilities collected with different time lags):
\begin{equation}
\label{eqn:design}
P=\left[ \begin{array}{cccc}
p_{26h}^{12} & p_{26h}^{13} & \cdots & p_{26h}^{n}\\
p_{28h}^{11} & p_{28h}^{12} & \cdots & p_{28h}^{n-1}\\
\vdots & \vdots& \ddots& \vdots\\
p_{48h}^{1} & p_{48h}^{2} & \cdots & p_{48h}^{n-11}
\end{array} 
\right ]
\end{equation}

$L$ denotes the ground truth (i.e., the CCCDF of $Dst$), and $W$ is the sought after weight vector. $W$ is initially set to a constant vector, and will be optimized by a least-square (LS) method:

\begin{equation}
\label{eqn:weight}
W=[\begin{array}{cccc}
     w_{26h} & w_{28h} & \cdots & w_{48h}\\
\end{array}]
\end{equation}

In this study, we develop a customized Elastic-net-aided LS method to optimize the weights $W$, based on the elastic-net regularization scheme proposed by Zou and Hastiea \cite{zou2005regularization}.  The residuals are calculated as: 

\begin{equation}
\label{eqn:V}
V=L - P\cdot W 
\end{equation}

The cost function, or so-called normal function in LS, is defined as:

\begin{equation}
\label{eqn:cost function}
F = V^TQV+r_1\sum{W}+r_2\sum{W^2}
\end{equation}

The first term in $F$ is a classic weighted cost function where $V^T$ is the transpose matrix of V. The second and third terms are Elastic net regularization factors that linearly combines the L1 and L2 penalties of the lasso and ridge methods. Based on the experiments on the storms, here we set $r_1=0.05$, and $r_2=0.95$. Finally, $Q$ denotes the `weight', i.e.,  diagonal matrix with positive and negative samples as defined in Eqn. \ref{eqn:Q}.

\begin{equation}
\label{eqn:Q}
q_{pos} = \frac{n_{all}}{n_{pos}},   q_{neg} = \frac{n_{all}}{n_{neg}}
\end{equation}

The function $F$ reaches a minimum when the partial derivative of $F$ with respect to $W$ equals zero:
\begin{equation}
\label{eqn:derivative}
\frac{\partial F}{\partial W} = -2V^TQP+\Vec{r}_1+2\Vec{r}_2{W}=0,
\end{equation}

where $\Vec{r}_1$ is a $12\times 1$ vector and $\Vec{r}_2$ is a $12\times 12$ unit vector times scale $r_2$.

or 

\begin{equation}
\label{eqn:derivative2}
P^TQV=\frac{1}{2}\Vec{r}_1+\Vec{r}_2{W}
\end{equation}

Multiplying $P^TQ$ with Eqn. \ref{eqn:V}, one has

\begin{equation}
\label{eqn:ATPA}
P^TQV=P^TQL - P^TQP\cdot W 
\end{equation}

Setting Eqn. \ref{eqn:derivative2} into Eqn. \ref{eqn:ATPA}, one has

\begin{equation}
\label{eqn:ATPA2}
\frac{1}{2}\Vec{r}_1+\Vec{r}_2{W}=P^TQL - P^TQP\cdot W 
\end{equation}

Finally,

\begin{equation}
\label{eqn:W}
W= (P^TQP+\Vec{r}_2)^{-1}(P^TQL-\frac{1}{2}\Vec{r}_1)
\end{equation}

\section{Results}
\label{sec:results}

In this section we show the results of our model in terms of the metrics TSS and MCC scores discussed in Section \ref{subsec:metrics}. It should be noted that, by using the leave-one-out technique, all metrics in this section are calculated based on the combination of all entries in the confusion matrix. 
The proposed ensemble approach is compared against three alternative approaches: support vector regression (SVR) which is a non-linear ensemble method \cite{awad2015support}, a method where ensemble members are simply averaged (equal weight, denoted as `Constant' in Table \ref{tab:methods}), and a single individual prediction 24 hrs ahead (no ensemble, so-called `Single' in Table \ref{tab:methods}). Table \ref{tab:methods} shows the TSS and MCC for the four methods. One can notice that the LS ensemble method significantly outperforms the single individual prediction model. Moreover, although SVR yields a large TNR, the FPR is also large, resulting in low values for both TSS and MCC.

\begin{table}[!htbp]
\centering
\caption{Accuracy of the proposed ensemble model with LS, SVR and constant weights, together with single-target CNN model. }
\begin{tabular}{c | c c | c c c c c}
  \hline
  Ensemble Method & TSS & MCC & TP & FP & TN & FN\\
  \hline
  LS & \textbf{0.62} & \textbf{0.37} & 57 & 209 & 1738 & 21 \\
  SVR & 0.15 & 0.18 & 13 & 37 & 1910  & 65 \\
  Constant & 0.32 & 0.23 & 31 & 135 & 1812 & 47 \\
  \hline
  Single & 0.28 & 0.17 & 103 & 544 & 2097 & 111 \\
\end{tabular}
\label{tab:methods}
\end{table}

As anticipated in Section \ref{sec:introduction}, the goal of this work is not just to provide a binary classification, but rather to estimate the probability of exceeding predefined thresholds. The LS-ensemble method has been trained and tested for various $Dst$ thresholds (Table \ref{tab:threshold}), different forecast duration (see Table \ref{tab:delay}) and  larger time resolution (Table \ref{tab:res}).  Table \ref{tab:threshold} demonstrates that the model performs best when the $Dst$ threshold is -100 nT. Corresponding TSS and MCC are 0.62 and 0.37, respectively. Both TSS and MCC decrease when the threshold is set to -50 nT. Although TSS performs well with a threshold of -200 nT, the corresponding MCC decreases significantly because of the imbalanced labels since very few `very strong' storms ($\le$ -200 nT) occurred during 1999-2009. This implies that the proposed method may identify strong storms ($\le$ -100 nT) better than mild storms ($\le$ -50 nT). Table \ref{tab:delay} shows how the performance of the model gets worse with a longer lead-time. From Table \ref{tab:res}, we can see that the model trained from 2-hrs samples outperforms the model trained from 6-hrs samples. This is because a 6-hrs samples traninig set is composed of fewer samples overall. Therefore, the model could be less robust with a larger cadence. Finally, we show in Table \ref{tab:wmet} the modified TSS and MCC scores proposed in \citeA{guastavino2021bad} that further improve the accuracy of the method.

\begin{figure}
  \centering
  \includegraphics[width=0.8\textwidth]{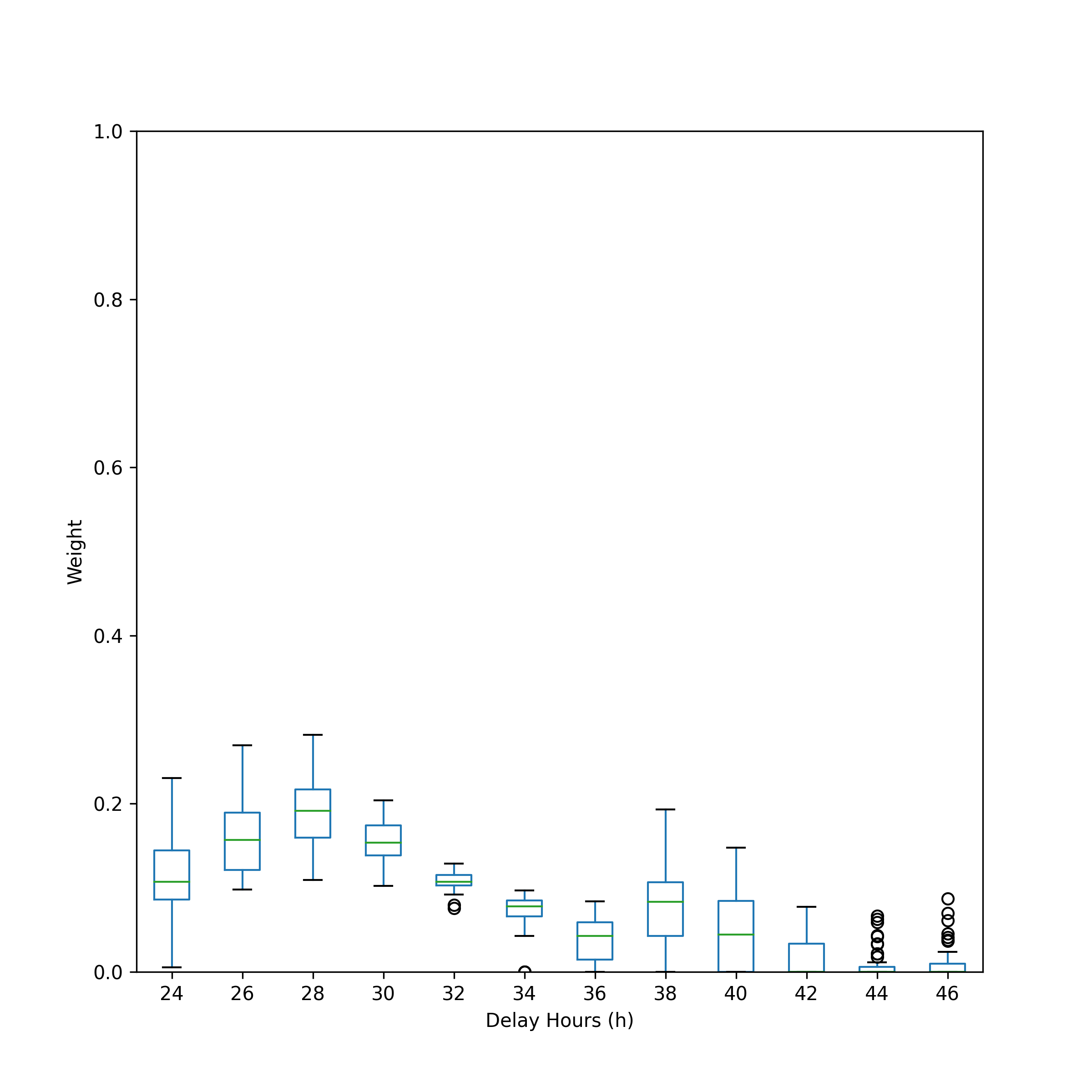}
  \caption{Boxplot of weights of 12 ensemble predictions. Green lines are median value for each weight, and black dots are anomalies. The upper and lower boundary of blue lines are the maximum and minimum. The upper and lower boundary of blue lines are the first and third quartile within 51 storm events.}
  \label{fig:weight}
\end{figure}

A statistic analysis of weights from all of the 51 sub-models is plotted in Fig. \ref{fig:weight}. Green horizontal lines denote the mean weights from all 51 sub-models. Blue bars represent the uncertainty range between the first and the third quartile of the distributions, and black dots are distribution outliers. Fig. \ref{fig:weight} shows that the predictions at 24-38 delay hours have the largest contributions to the final probability. The weights decrease with the increase of delay hours. It is interesting that the contribution of predictions during 42 to 46 delay hours are essentially negligible. The error bars and presence of outliers also imply that the weights vary slightly according to different storms. This may be improved by having more representative storm events.

\begin{table}[!htbp]
\centering
\caption{Accuracy of the developed model with 24 hrs ahead predictions based on different $Dst$ threshold, i.e., -50nT, -100nT and -200nT}
\begin{tabular}{c | c c | c c c c c}
  \hline
  $Dst$ threshold & TSS & MCC & TP & FP & TN & FN\\
  \hline
  -50 nT & 0.29 & 0.26 & 215 & 350 & 1254 & 206 \\
  -100 nT & \textbf{0.62} & \textbf{0.37} & 57 & 209 & 1738 & 21 \\
  -200 nT & 0.56 & 0.14 & 3 & 80 & 1940 & 2 \\
\end{tabular}
\label{tab:threshold}
\end{table}


\begin{table}[!htbp]
\centering
\caption{Accuracy of the developed model with 1, 2 and 3 days ahead, when the $Dst$ threshold is -100 nT. }
\begin{tabular}{c | c c | c c c c c}
  \hline
  Forecast (Days) & TSS & MCC & TP & FP & TN & FN\\
  \hline
  1 & \textbf{0.62} & \textbf{0.37} & 57 & 209 & 1738 & 21 \\
  2 & 0.34 & 0.30 & 40 & 93 & 1807  & 63 \\
  3 & 0.27 & 0.17 & 44 & 248 & 1578 & 65 \\
\end{tabular}
\label{tab:delay}
\end{table}

\begin{table}[!htbp]
\centering
\caption{Accuracy of the developed model with different time resolution. }
\begin{tabular}{c | c c | c c c c}
  \hline
  time resolution & TSS & MCC & TP & FP & TN & FN \\
  \hline
  2 & \textbf{0.62} & \textbf{0.37} & 57 & 209 & 1738 & 21 \\
  6 & 0.42 & 0.25 & 9 & 47 & 555 & 9 \\
\end{tabular}
\label{tab:res}
\end{table}

\begin{table}[!htbp]
\centering
\caption{Weighted metrics of the developed model with different time resolution.}
\begin{tabular}{c | c c | c c c c}
  \hline
  time res & wTSS & wMCC & wTP & wFP & wTN & wFN\\
  \hline
  2 & \textbf{0.68} & \textbf{0.47} & 67 & 133 & 1407 & 20 \\
  6 & 0.59 & 0.42 & 41 & 82 & 510 & 15 \\
\end{tabular}
\label{tab:wmet}
\end{table}

\subsection{Storm Case Study}
\label{subsec:storm_case}

In this section we would like to investigate several typical storm cases. 

\begin{figure}
  \centering
  \includegraphics[width=1.\textwidth]{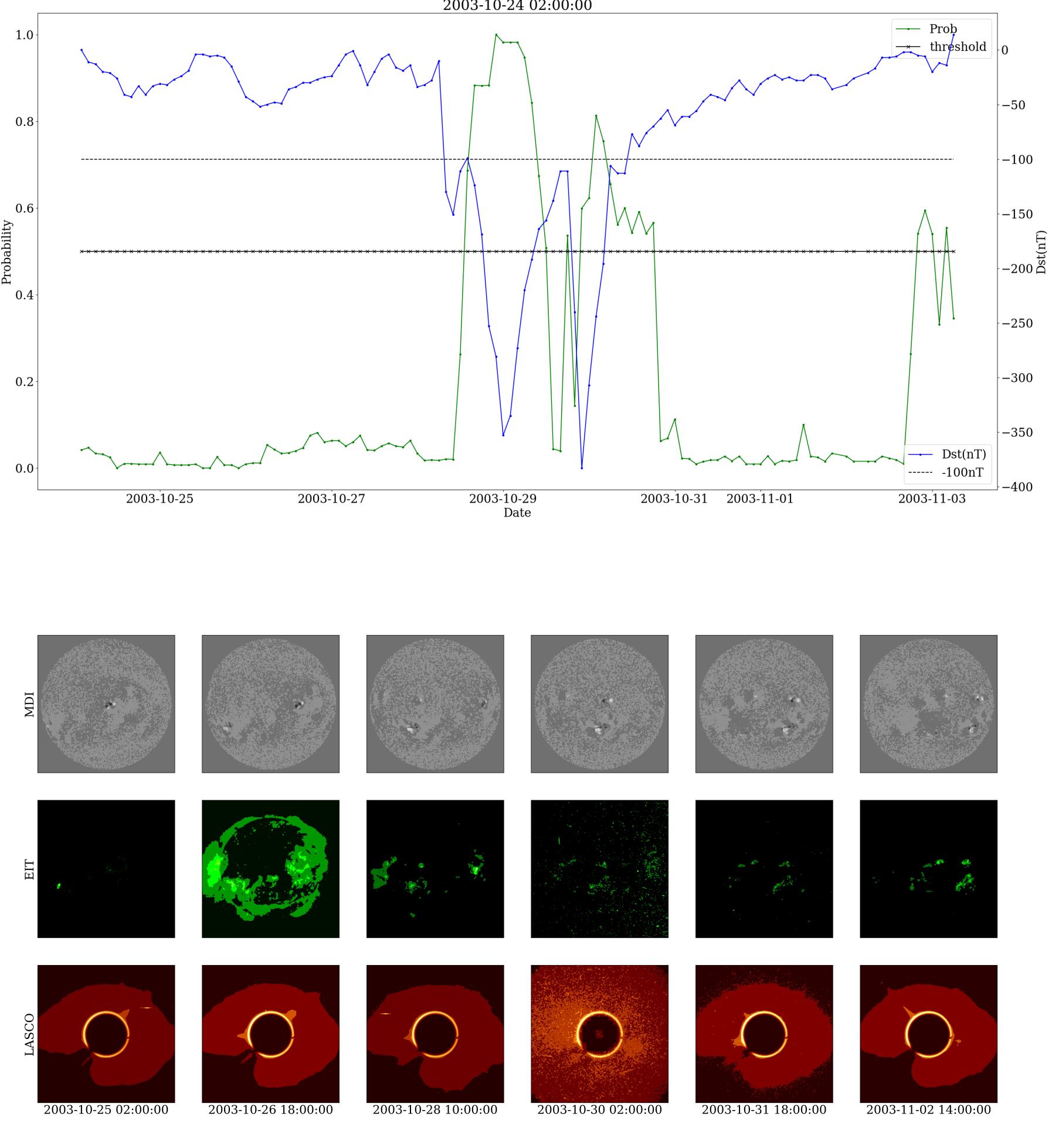}
  \caption{Probabilities generated from the developed model during 2003-10-25 and 2003-11-05, together with the corresponding $Dst$ and the thresholds for both probabilities and $Dst$ during this storm cases. X axis is Date. Left and right Y axes are probability (shown by green line) and $Dst$ (shown by blue line) respectively. Cross line is when the probability equals to 0.5, and dash line is when $Dst$ equals to -100 nT. Corresponding images are plotted in the bottom panel. From top to bottom, they are for MDI, EIT-195 and LASCO C2.}
  \label{fig:Halloween}
\end{figure}

The Halloween storm, caused by a CME, from 2003-10-25 to 2003-11-05 is selected for a case study. This is the biggest storm in the past 20 years. Figure \ref{fig:Halloween} displays the 24-hour ahead $Dst$ probabilities of the developed model (green line, left vertical axis) and the corresponding $Dst$ (blue line, right vertical axis) during the Halloween storm. Black cross and dotted lines are the threshold of probability and $Dst$ index, respectively. 

The predicted probabilities can be converted to a binary format with an optimal threshold re-calibrated by training samples. This threshold is rescaled back to 0.5 in Fig. \ref{fig:Halloween}. Two peaks at the midnight October 29 \& 30 can be well captured by these predicted probabilities. The time shift between the peak of real $Dst$ and the peak of the predicted probability is no more than 4 hours. This implies that those strong storm can predicted very well by the proposed model. 

\begin{figure}
  \centering
  \includegraphics[width=1.\textwidth]{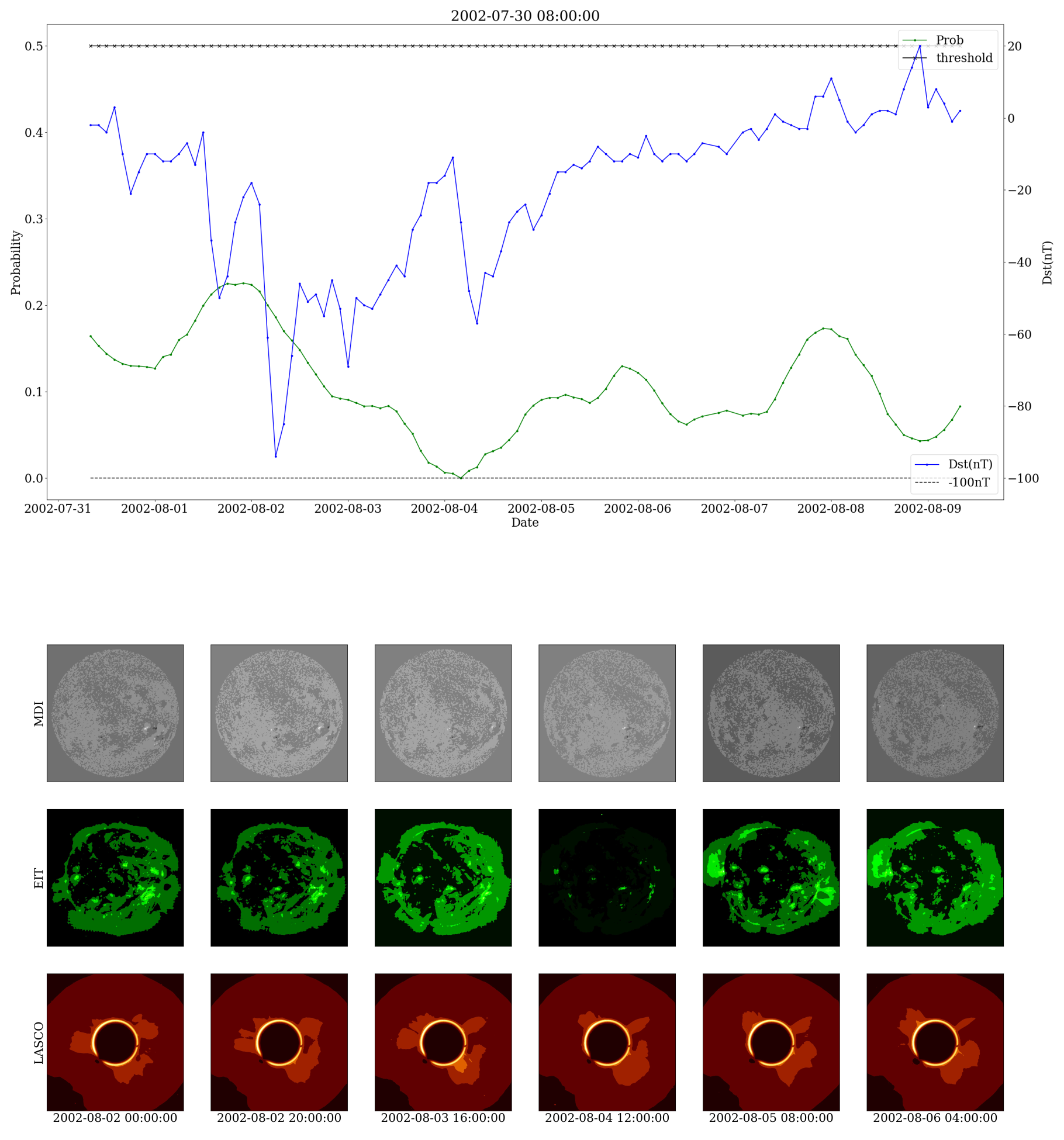}  
  \caption{Similar to Fig. \ref{fig:Halloween}. An example during an non-Earth-direct CME during 2002-07-31 and 2002-08-02.}
  \label{fig:nonearth}
\end{figure}

Fig. \ref{fig:Halloween} also indicates that the proposed model can forecast well those storms caused by CME because they stand out very clearly in SoHO images. However, a good portion of solar CMEs are non-Earth oriented. The ability of this model to identify the geo-effectivenes (or lack thereof) of non-Earth-directed CMEs is also assessed. Twenty of them occurring in the period 2000-2003 are selected for validation. The TSS and MCC of the prediction based on the developed model during those non-Earth-directed CME periods are 0.94 and 0.46 respectively. The TP, FP, TN and FN values are 14, 50, 709 and 0. An example is shown in Fig. \ref{fig:nonearth}. A strong CME occurred around 2002-07-30, but the corresponding $Dst$ did not reach -100 nT. The predicted $Dst$ probability increases but not as significantly as for the Earth-directed CMEs. This suggests that the proposed model may be able to distinguish non-Earth-directed CMEs, and assign lower $Dst$ probabilities to them. 
Similar plots of probabilities for all the other storm events used in this study are included as supplementary information. 
\section{Summary \& Outlook}
\label{sec:summary-outlook}

We have developed a LS-based class-balanced ensemble CNN model that estimates the probability of $Dst$ exceeding a given threshold 1 day ahead based on SoHO images. 51 selected storm events were chosen during a long‐span historical data set ($\sim$ 16 years), between 1996-05-01 and 2011-04-20. The proposed model can predict the probability that $Dst<-100 $nT 24 hours ahead with a TSS of 0.62 and MCC of 0.37. The weighted TSS and MCC from \citeA{guastavino2021bad} are 0.68 and 0.47.

One of the crucial points of this work is that it combines a LS ensemble method with a CNN algorithm for a probability prediction. A customized class-balanced mean square error is developed as the cost function of the proposed CNN model. After the CNN model is developed, a LS method is developed to estimate the weights of the predictions from the proposed multi-target CNNs. Eventually a final probability value can be calculated by the optimized weights and CNN predictions. Binary classification of each event is then determined by a threshold re-calibrated on the other 50 storms used for training. 

We have shown that this proposed model provides good skills for predicting $Dst$ 1-day-ahead during strong storm periods. The proposed model can also forecast $Dst$ probability even within a non-Earth-direct CME period. This model will extend the prediction lead time of most of the current $Dst$ empirical prediction models. The performance metrics that we have analyzed are the confusion matrix, TSS, MCC, and corresponding weighted scores from \cite{guastavino2021bad}. Finally, we have discussed a strong storm case from 2003-10-25 to 2003-11-05. $Dst$ peaks can be well captured by the developed model.

A possible weakness of this model is that the weights are static, although the time lags from different events should not be the same \cite{chandorkar2019dynamic}. As a next step, we plan to take into account dynamic weights, e.g. by applying an online/dynamic ensemble method, such as \cite{monteleoni2011tracking}.  Moreover, storms that result from CMEs or high speed streamers are based on different physical mechanisms. The former mostly occur during high solar activity period, while the latter are seen more often during low solar activity period. Including a solar activity index, such as $F_{10.7}$ into consideration to first classify those samples in order to train the model more precisely, will be also experimented in the future. 

\acknowledgments

This project has been developed in the framework of the European project AIDA. The AIDA project has received funding from the European Union’s Horizon 2020 Research and Innovation programme under grant agreement No 776262. 
EC is partially funded by the National Aeronautics and Space Administration under grants 80NSSC20K1580, 80NSSC20K1275, 80NSSC21K1555. We thank OMNIWeb for providing the $Dst$ data (https://omniweb.gsfc.nasa.gov/), the NASA Solar Data Analysis Center’s (SDAC) Virtual Solar Observatory (VSO) (\url{https://sdac.virtualsolar.org/cgi/search}) and Stanford University’s Joint Science Operation Center (JSOC) \url{http://jsoc.stanford.edu/MDI/MDI_Magnetograms.html} for SoHO images. All the results and codes have been made available as a Zenodo repository in 10.5281/zenodo.6385325. Future updates can be found on https://github.com/HuanWinter/Dst\_SoHO and https://ml-space-weather.github.io/projects.html (under construction)

\bibliography{main}

\newpage

\end{document}